\documentclass[aps,pre,twocolumn,floatfix]{revtex4}
\usepackage[]{natbib}

\usepackage{graphicx}
\usepackage{amsmath}
\usepackage{amsfonts}
\usepackage{amssymb}
\usepackage{bm}

\begin{document}

\title{On the synchrotron emission in kinetic simulations of runaway electrons in magnetic confinement fusion plasmas\footnote{This manuscript has been authored by UT-Battelle, LLC under Contract No. DE-AC05-00OR22725 with the U.S. Department of Energy. The United States Government retains and the publisher, by accepting the article for publication, acknowledges that the United States Government retains a non-exclusive, paid-up, irrevocable, worldwide license to publish or reproduce the published form of this manuscript, or allow others to do so, for United States Government purposes. The Department of Energy will provide public access to these results of federally sponsored research in accordance with the DOE Public Access Plan (http://energy.gov/downloads/doe-public-access-plan).}}

\author{L. Carbajal}
\author{D. del-Castillo-Negrete}
\affiliation{Oak Ridge National Laboratory \\ Oak Ridge, Tennessee 37831-8071, USA}

\begin{abstract}

Developing avoidance or mitigation strategies of runaway electrons (RE) in magnetic confinement fusion (MCF) plasmas is of crucial importance for the safe operation of ITER. In order to develop these strategies, an accurate diagnostic capability that allows good estimates of the RE distribution function in these plasmas is needed.
Synchrotron radiation (SR) of RE in MCF, besides of being one of the main damping mechanisms for RE in the high energy relativistic regime, is routinely used in current MCF experiments to infer the parameters of RE energy and pitch angle distribution functions.
In the present paper we address the long standing question about what are the relationships between different runaway electrons distribution functions and their corresponding synchrotron emission simultaneously including: full-orbit effects, information of the spectral and angular distribution of SR of each electron, and basic geometric optics of a camera.
We study the spatial distribution of the SR on the poloidal plane, and the statistical properties of the expected value of the synchrotron spectra of runaway electrons. 
We observe a strong dependence of the synchrotron emission measured by the camera on the pitch angle distribution of runaways, namely we find that crescent shapes of the spatial distribution of the SR as measured by the camera relate to RE distributions with small pitch angles, while ellipse shapes relate to distributions of runaways with larger the pitch angles. A weak dependence of the synchrotron emission measured by the camera with the RE energy, value of the $q$-profile at the edge, and the chosen range of wavelengths is observed. Furthermore, we find that oversimplifying the angular dependence of the SR changes the shape of the synchrotron spectra, and overestimates its amplitude by approximately 20 times for avalanching runaways and by approximately 60 times for mono-energetic distributions of runaways.

\end{abstract}

\maketitle

\section{Introduction}

Runaway electrons (RE), thermal electrons accelerated to relativistic energies during the rapid termination of a magnetic confinement fusion (MCF) plasma, pose a threat to ITER if they are not avoided or mitigated before they hit the wall, causing damage to plasma facing components \cite{Hender2007,Boozer2015}.
Various strategies to avoid or mitigate RE in MCF plasmas have been proposed, e.g. using resonant magnetic perturbations (RMPs) to deconfine RE before they reach high energies~\cite{Papp2011}, or using either massive gas injection (MGI) or shattered pellet injection (SPI) of high $Z$ impurities to slow down RE through collisional drag and by enhancing synchrotron radiation losses of RE through pitch angle scattering driven by collisions~\cite{Lehnen2011,Hollmann2015,Commaux2016}.
An accurate diagnostic capability that allows good estimates of the RE parameters in MCF plasmas is needed to gain a better understanding of the underlying physics of the RE dynamics, as well as  to guide the development of the avoidance and mitigation strategies.

Synchrotron radiation (SR) of RE in MCF plasmas is important for two reasons: SR is one of the main damping mechanisms for RE in the high energy relativistic regime, limiting the maximum energy that RE can reach during a disruption~\cite{Martin-Solis1998,Andersson2001}, and substantially reducing the runaway electron rate for weak (near critical) $E$ fields~\cite{Stahl2015}.
On the other hand, SR is routinely measured in current MCF experiments to infer the RE energy and pitch angle distribution function. The latter is done by interpreting the measured SR spectra and geometric features of the radiation spatial patterns seen by the visible and infrared cameras~\cite{Finken1990,Jaspers1995,Entrop1999,Jaspers2001,Kudyakov2008,DeAngelis2010,Shi2010}. 
Most recently, SR was used to infer the characteristic energy and pitch angle of RE in the DIII-D tokamak when MGI was used as the mitigation mechanism ~\cite{Yu2013,Hollmann2013}. 
In these plasmas the RE parameters were obtained by fitting the measured SR spectra with the single-particle spectrum, that is, the spectrum of a single electron calculated with a single energy and pitch angle, and using characteristic parameters of the plasma as measured at the magnetic axis.
In Ref.~\cite{Stahl2013} the authors showed that using single-particle spectra overestimates the SR by orders of magnitude and this can be misleading when inferring the RE parameters, and in general one should iteratively use the SR spectrum of a guess distribution function for the RE until the best fit to the experimental data is found. 
This overestimation, that depending on the wavelength can reach several orders of magnitude, is due to assuming that all runaways emit as much synchrotron radiation as the most strongly emitting particle in the actual runaway distribution function.
Later, the study of Ref.~\cite{Landreman2014} went a step further by solving the Fokker-Planck equation to obtain the RE distribution function for a given set of plasma parameters, and then calculating the corresponding SR spectrum. Again, the authors found that the single-particle spectra can be misleading when used to infer the RE parameters of more realistic RE distribution functions.
Importantly, the above studies did not include any information of the electrons' orbits, thus ignoring confinement and collisionless pitch angle scattering that can substantially modify the SR spectra \cite{Carbajal2017}.

In the present paper we address the long standing question about what are the relationships between different  runaway electrons distribution functions and their corresponding synchrotron emission simultaneously including: full-orbit effects, information of the spectral and angular distribution of synchrotron radiation of each electron, and basic geometric optics of a camera.
We follow the full-orbit dynamics of ensembles of runaway electrons in DIII-D-like plasmas using the new Kinetic Orbit Runaway electrons Code (KORC) to generate synthetic data to calculate different aspects of their synchrotron radiation.
First, we use mono-energy and mono-pitch angle distributions of runaways as initial conditions in our simulations to study the spatial distribution of the synchrotron radiation on the poloidal plane, and the statistical properties of the expected value of the synchrotron spectra of runaway electrons.
Then, we find relations between the runaway electrons' parameters and both the spatial distribution of the synchrotron emission and the synchrotron spectra as observed by a camera placed at the outer midplane plasma.
Finally, we use these results to interpret the synchrotron emission for an avalanche RE distribution function.
In our simulations we observe a strong dependence of the spatial distribution of the radiation on the pitch angle distribution of the runaways. Also, we find that the synchrotron spectra is very sensitive to oversimplifications of the angular distribution of the synchrotron radiation, dramatically changing its shape and amplitude.

The rest of the paper is organised as follows: in Sec.~\ref{theory} we present a brief summary of the theory of the synchrotron radiation used throughout the paper. In Sec.~\ref{sim_setup} we describe the parameters used in our simulations. In Sec.~\ref{results} we present the study of the relationship between various distribution functions of runaway electrons and their synchrotron emission on the poloidal plane and as measured by a camera. Finally, in Sec.~\ref{conclusions} we summarise our results and discuss their implications in the interpretation of experimental data. Details on the synthetic camera diagnostic are provided in the appendix.

\section{Synchrotron radiation theory}
\label{theory}

In our full-orbit simulations of runaway electrons we calculate the total radiated power, the synchrotron radiation spectra, and the spectral and angular distribution of the radiation.
The total instantaneous synchrotron radiated power of a relativistic electron moving in an arbitrary orbit with velocity $v$ is given by:

\begin{equation}
P_{T}= \frac{e^2}{6\pi \epsilon_0 c^3} \gamma^4 v^4 \kappa^2,
\label{Ptot}
\end{equation}

\noindent
where $\kappa$ is the instantaneous curvature of the electron orbit, $\gamma=1/\sqrt{1 - v^2/c^2}$ is the relativistic factor, $e$ is the magnitude of the electron charge, $c$ is the speed of light, and $\epsilon_0$ is the vacuum permittivity. For a relativistic electron moving in an electric $\bm{E}$ and magnetic $\bm{B}$ field, the instantaneous curvature $\kappa$ is given by:

\begin{equation}
\kappa = \frac{e}{\gamma m_e v^3} |\bm{v}\times \left( \bm{E} + \bm{v}\times\bm{B} \right)| \, .
\label{kappa}
\end{equation}

In the case when $\bm{E} \ll \bm{v}\times \bm{B}$, the curvature can be approximated as $\kappa \approx eB \sin{\theta}/\gamma m_e v$, where $\theta$ is the pitch angle of the electron, that is, the angle between the vectors $\bm{v}$ and $\bm{B}$.

The spectral distribution of the synchrotron radiation of relativistic electrons is given by \cite{Schwinger1949}:

\begin{equation}
P_R(\lambda) = \frac{1}{\sqrt{3}} \frac{ce^2}{\epsilon_0 \lambda^3} \left( \frac{mc^2}{\mathcal{E}}\right)^2\int_{\lambda_c/\lambda}^\infty K_{5/3}(\eta)d\eta
\label{P_lambda}
\end{equation}

\noindent
Here, $\mathcal{E}=\gamma m_e c^2$ is the relativistic electron's energy, $K_{5/3}(\eta)$ is the modified Bessel function of the second kind of order $5/3$, and $\lambda$ is the wavelength at which the electron is radiating. The critical wavelength $\lambda_c = 4\pi/(3\kappa \gamma^3)$ is the wavelength characterizing $P_R(\lambda)$, dividing the spectra into two parts of equal radiated power, that is, half the total power is radiated at wavelengths $\lambda > \lambda_c$, and the rest is radiated at wavelengths $\lambda < \lambda_c$ \cite{Mobilio2015}. We should note that Eq.~(\ref{P_lambda}) is completely general and can be used for calculating the synchrotron spectrum of a relativistic electron moving in an arbitrary orbit with radius of curvature $1/\kappa$.  In Ref.~\cite{Pankratov1999} an approximate expression for the spectral distribution of the synchrotron radiation of runaway electrons with small pitch-angle in tokamaks was derived, and used in Refs.~\cite{Jaspers2001,Yu2013,Stahl2013}.

The most detailed level of description for the synchrotron radiation emitted by a relativistic electron is given by its spectral and full angular distribution, which in the case when the angle between the direction of emission and motion is small is given by:
	
\begin{eqnarray}
P_R(\lambda,\psi,\chi) & = &  \frac{c e^2}{\sqrt{3}\epsilon_0 \kappa \lambda^4\gamma^4} \left( 1 + \gamma^2 \psi^2 \right)^2 \times \nonumber \\
& & \left\lbrace \frac{\gamma^2 \psi^2}{1 + \gamma^2 \psi^2} K_{1/3}(\xi) \cos\left[\frac{3}{2}\xi\left( z + \frac{1}{3}z^3\right) \right] \right. \nonumber \\
& & \left. -\frac{1}{2}K_{1/3}(\xi)\left( 1 + z^2 \right)\cos\left[ \frac{3}{2}\xi\left( z + \frac{1}{3}z^3\right) \right] \right. \nonumber \\
& & \left. +  K_{2/3}(\xi)z\sin\left[\frac{3}{2}\xi\left( z + \frac{1}{3}z^3\right) \right] \right\rbrace \, ,
\label{P_ang}
\end{eqnarray}

\noindent
where $\xi = 2\pi \left( 1/\gamma^2 + \psi^2 \right)^{3/2}/3\lambda\kappa$, $z = \gamma \chi/\sqrt{1 + \gamma^2 \psi^2}$, $\psi$ is the angle between the direction of emission $\hat{\bm{n}}$ and the instantaneous orbital plane containing the tangent and normal vectors, that is, $\psi$ is the complementary angle to the angle between $\hat{\bm{n}}$ and the binormal vector defined below; $\chi$ is the angle between the projection of $\hat{\bm{n}}$ on the instantaneous orbital plane and the instantaneous direction of motion $\bm{v}$, respectively. The unit vector defining the direction of emission $\hat{\bm{n}}$ points from the electron's position towards where an observer measuring the radiation is. 
It is worth mentioning that Eq.~(\ref{P_ang}) is obtained when going from Eq.~(II.31) to Eq.~(II.32) of Ref.~\cite{Schwinger1949}.
In Eq.~(\ref{P_ang}) it is assumed that the synchrotron radiation is emitted mainly along $\bm{v}$, that is, small $\psi$ and $\chi$. $K_{1/3}$ and $K_{2/3}$ are the modified Bessel functions of second kind of order $1/3$ and $2/3$, respectively.
The instantaneous orbital plane of the electron is uniquely determined by the tangent vector $\hat{\bm{\mathcal{T}}}$, which corresponds to the unit electron velocity $\hat{\bm{v}} = \bm{v}/v$, the normal vector $\hat{\bm{\mathcal{N}}}$, and the binormal vector $\hat{\bm{\mathcal{B}}} = \hat{\bm{\mathcal{T}}} \times \hat{\bm{\mathcal{N}}}$, which is perpendicular to the instantaneous orbital plane. For a relativistic electron moving in an arbitrary electric and magnetic field

\begin{equation}
\hat{\bm{\mathcal{B}}} = \frac{\bm{v}\times{\dot{\bm{v}}}}{|\bm{v}\times{\dot{\bm{v}}}|} = \frac{\bm{v}\times \left( \bm{E} +  \bm{v}\times \bm{B}\right)}{|\bm{v}\times \left( \bm{E} +  \bm{v}\times \bm{B}\right)|}\, .
\end{equation}


\noindent
$P_R(\lambda,\psi,\chi)$ in Eq.~(\ref{P_ang}) decreases exponentially as a function of $\psi$ through the function $\xi$, this due to $K_{1/3}(\xi)$ and $K_{2/3}(\xi)$. On the other hand Eq.~(\ref{P_ang}) shows oscillations as a function of $\chi$ through the function $z$, and can become negative for large values of $\chi$ or $\psi$.
In order to make an efficient search on the $\psi\chi$-plane where $P_R(\lambda,\psi,\chi)$ is positive and thus physically meaningful, we restrict our search to to a rectangular domain containing the region of validity defined by the values \cite{jackson2007classical}:

\begin{equation}
\psi_c = \left( \frac{3\kappa \lambda}{4\pi} \right)^{1/3} \, ,
\label{psi_c}
\end{equation}

\noindent
and $\chi_c$, which is a solution of the equation:

\begin{equation}
\frac{\gamma^3}{3}\chi_c^3 + \gamma \chi_c - \frac{\pi}{3\xi} = 0\, .
\label{chi_c}
\end{equation}

\begin{figure}[ht!]
\begin{center}
\includegraphics[scale=0.475]{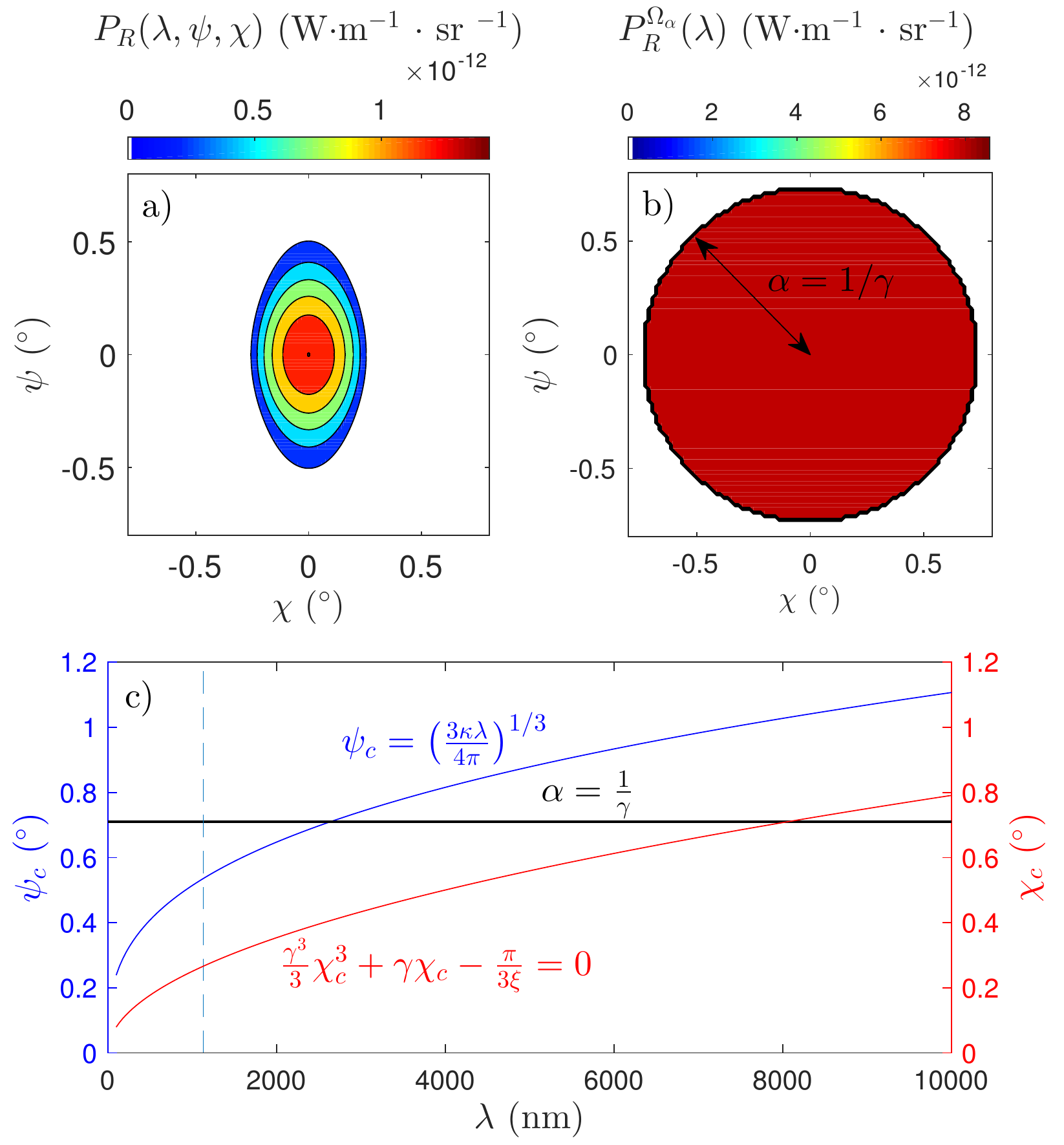}
\end{center}
\caption{Filled contours of the angular distribution of the synchrotron radiation emitted by a relativistic electron with $\mathcal{E} = 30$ MeV, and $\theta_0 = 10^\circ$ at the HFS of an DIII-D-like magnetic field. The magnetic field at the magnetic axis is $B_0 = 2.19$ T and the safety factor at the magnetic axis and the edge are $q=1$, and $q=3$, respectively. Panel a): full spectral and angular distribution $P_R(\lambda,\psi,\chi)$ of Eq.~(\ref{P_ang}). Panel b): simplified model for the angular distribution $P_{R}^{\Omega_\alpha}(\lambda)$. Panel c): upper limits of the angles $\psi$ and $\chi$ of the spectral and angular distribution of Eq.~(\ref{P_ang}) as a function of the wavelength. The horizontal black line shows the values of $\psi_c$ and $\chi_c$ for the simplified model for the angular distribution $P_{R}^{\Omega_\alpha}(\lambda)$. The dashed, vertical line shows the wavelength at which panels a) and b) are computed.}
\label{P_ang_contours}
\end{figure}

Fig.~\ref{P_ang_contours}(a) shows an example of $P_R(\lambda,\psi,\chi)$ in the domain defined by $\psi_c$ and $\chi_c$ for a relativistic electron with energy $\mathcal{E} = 30$ MeV and pitch angle $\theta_0 = 10^\circ$ at the high-field side (HFS) of a DIII-D-like magnetic field. 
From this figure we observe that the synchrotron radiation is emitted within an ellipse in the $\psi$ and $\chi$ plane with major and minor radii bounded by $\psi_c$ and $\chi_c$. This means that the radiation is emitted within an elliptic cone with its axis along $\hat{\bm{v}}$.

Previous studies where synchrotron emission has been used to diagnose runaway electrons \cite{Finken1990,Jaspers1995,Yu2013,Wongrach2014} have simplified the synchrotron angular distribution to either a $\delta$ function in space, that is, $P_{R}^\delta(\lambda) = P_R(\lambda)\cdot \delta (\psi)\cdot \delta (\chi)$, or to a circular cone with ``natural aperture'' $\alpha = 1/\gamma$, that is, $P_{R}^{\Omega_\alpha}(\lambda)  =  P_R(\lambda)/\Omega_\alpha$ for $\psi^2 + \chi^2 \leq \alpha^2$, and $P_{R}^{\Omega_\alpha}(\lambda) =0$ otherwise.
In the former case, all the radiation $P_R(\lambda)$ of Eq.~(\ref{P_lambda}) is emitted along the velocity of the particle, while in the latter case $P_R(\lambda)$ is allowed to ``spread'' uniformly within the solid angle subtended by $\alpha$, that is, $\Omega_\alpha = 2\pi [1 - \cos(\alpha) ]$. Throughout this paper, we will refer to $P_{R}^{\Omega_\alpha}(\lambda)$ as the simplified model for the angular distribution of the synchrotron radiation. In Fig.~\ref{P_ang_contours}(b) we show the corresponding $P_{R}^{\Omega_\alpha}(\lambda)$ for the same values of Fig.~\ref{P_ang_contours}(a). 

\section{Simulations setup}
\label{sim_setup}

In order to study the relationship between the runaway electron distribution functions and different aspects of their synchrotron radiation we have used the new Kinetic Orbit Runaway electrons Code (KORC). This code is a parallel Fortran 95 code that follows large ensembles of runaway electrons in the full 6-D phase space. KORC efficiently exploits the shared memory computational systems by using the hybrid open MP + MPI paradigm for parallelisation, showing nearly ideal weak and strong scaling. KORC incorporates the Landau-Lifshitz formulation of the radiation reaction force \cite{Carbajal2017}, and Coulomb collisions of RE with the thermal plasma using the model of Ref.~\cite{Papp2011}.

In all the simulations reported in this paper we have used the analytical field of Ref.~\cite{Carbajal2017} with DIII-D-like plasma parameters where RE occur, that is, the magnitude of the magnetic field at the magnetic axis $B_0=2.1$ T, safety factor $q_0 = 1$ at the magnetic axis and $q_{edge} = 3$ at the plasma edge, which is located at $r_{edge} = 0.5$ m. The major radius of the plasma is $R_0 = 1.5$ m. The magnetic field model in toroidal coordinates is given by:

\begin{equation}
\bm{B}(r,\vartheta) = \frac{1}{1 + \eta \cos{\vartheta}} \left[ B_0 \hat{\bm{e}}_\zeta  + B_\vartheta(r) \hat{\bm{e}}_\vartheta \right].
\label{Eq3}
\end{equation}

\noindent
where $\eta = r/R_0$ is the aspect ratio, $B_\vartheta(r) = \eta B_0/q(r)$ is the poloidal magnetic field. The safety factor is

\begin{equation}
q(r) = q_0\left( 1 + \frac{r^2}{\varepsilon^2} \right) \, .
\label{Eq4}
\end{equation}

\noindent
The constant $\varepsilon$ is obtained from the values of $q_0$ and $q(r)$ at the plasma edge $r=r_{edge}$. The coordinates $(r,\vartheta, \zeta)$ are
defined as  $x= \left( R_0 + r \cos \vartheta \right ) \sin \zeta$, $y =\left( R_0 + r \cos \vartheta \right ) \cos \zeta$, and  $z =r \sin \vartheta$, where $(x,y,z)$ are the Cartesian coordinates. In these coordinates, $r$ denotes the minor radius, $\vartheta$ the poloidal angle, and $\zeta$ the toroidal angle. Note that in this right-handed toroidal coordinate system, the toroidal angle $\zeta$ rotates clockwise, that is, it is anti-parallel to the azimuthal angle, $\phi=\pi/2-\zeta$, of the standard cylindrical coordinate system.

Throughout the paper we use different distribution functions for the runaway electrons in the energy and pitch-angle space $f_{RE}(\mathcal{E},\theta)$. We use $5\times 10^6$ computational particles uniformly distributed in a torus as the spatial distribution of the different $f_{RE}(\mathcal{E},\theta)$, see for example Fig.~\ref{PR_poloidal_plane_30MeV}(c) or Fig.~\ref{camera_setup}(a). The major radii of the torus and the radii of the RE beam used for the spatial initial condition are specified in each section, and are chosen so that all the runaways remain confined during the simulation \cite{Carbajal2017}.
In our simulations the plasma current is anti-parallel to the toroidal electric field.
The simulation time $t_{sim} \sim 10 \ \mu$s is set so that the less energetic RE considered in our simulations undergo 30 poloidal turns. Because this time is much smaller than both the collisional time $\tau_{coll} = 4\pi \epsilon_0^2 m_e^2 c^3/(n_e e^4 \log{\Lambda})\sim 10 \ \mbox{ms}$ and the characteristic time for radiation losses $\tau_{R} = 6\pi\epsilon_0(m_ec)^3/(e^4B^2) \sim 1\ \mbox{s}$, we have turned off the radiation reaction force and collisions in KORC, so the energy is conserved in these simulations. This simulation time is observed to be enough for reaching a collisionless steady-state distribution function, that is, a time independent solution of the full orbit $f_{RE}(\mathcal{E},\theta)$.



\section{Full-orbit effects on synchrotron emission of various RE distribution functions}
\label{results}

In this section we study the collisionless pitch angle dispersion effects on the synchrotron radiation spectra emitted by various runaway electron distribution functions. 
Previous studies using a full-orbit description of RE in toroidal plasmas \cite{Liu2016,Wang2016,Carbajal2017} have shown that due to the variation of the magnetic field seen by runaways along their orbits, they experience collisionless pitch angle dispersion, even in the the case where collisions or synchrotron radiation losses are not included. Because the synchrotron radiation of each electron strongly depends on its pitch angle, it is expected that the resulting synchrotron emission of different ensembles of runaway electrons will show non-trivial changes with respect to the results inferred from distributions that do not take into account collisionless pitch angle dispersion effects. The aim of this sections is to study these changes in detail.

\subsection{Synchrotron emission of mono-energetic and mono-pitch angle RE distributions on the poloidal plane}
\label{mono-section_poloidal}

\begin{figure*}[ht!]
\begin{center}
\includegraphics[scale=0.55]{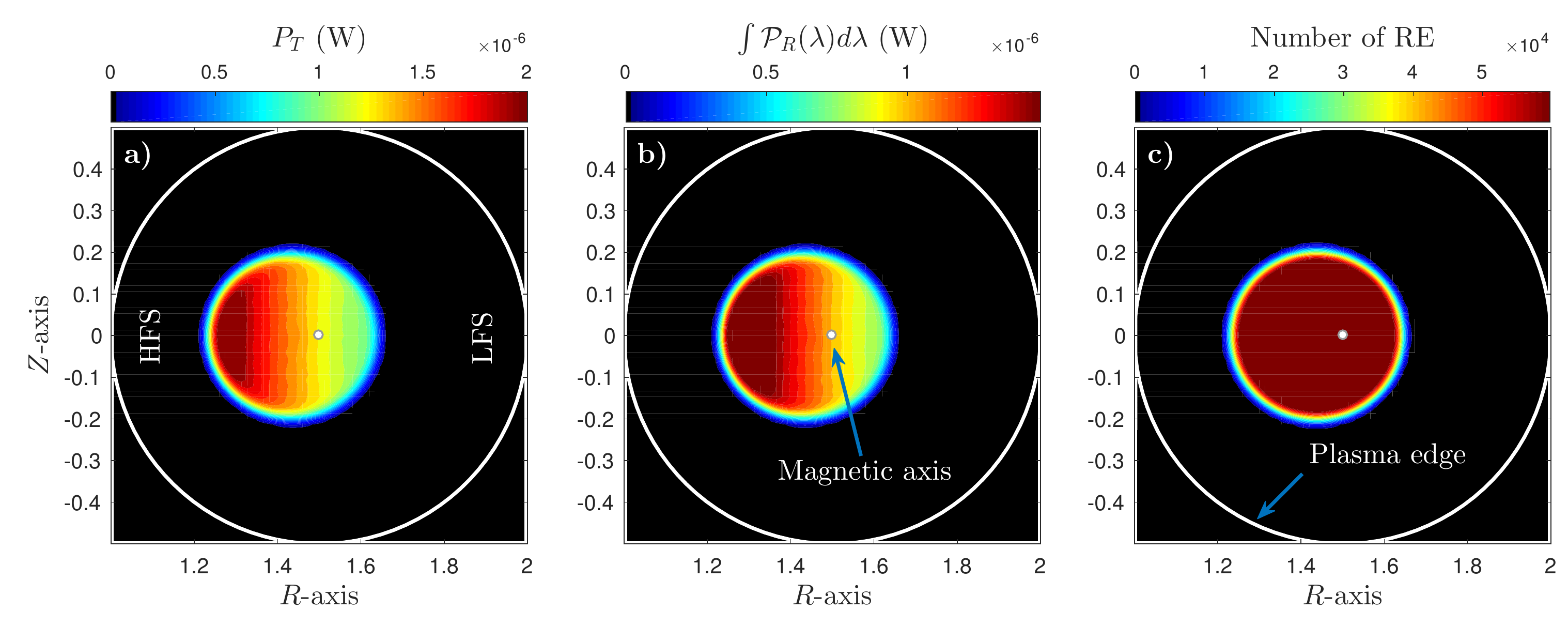}
\end{center}
\caption{Spatial distribution on the poloidal plane of the total and integrated synchrotron radiated power of a simulated ensemble of runaway electrons with initial $\mathcal{E}=30$ MeV and $\theta_0=10^\circ$. Panel a): spatial distribution of the total synchrotron radiated power $P_T$ of Eq.~(\ref{Ptot}). The radiation is more intense at the HFS and less intense at the LFS. An up-down symmetry is observed. Panel b): spatial distribution of the radiated power $P_R(\lambda)$ of Eq.~(\ref{P_lambda}) integrated over $\lambda\in (100,10000)$ nm. The same qualitative features of $P_T$ are observed. Panel c): spatial distribution of the full orbit RE distributions. These same features of $P_T$ and the integrated synchrotron radiation power are observed in all the other simulations of initially mono-energetic and mono-pitch angle distributions of runaway electrons. For producing these figures we computed the histograms of each quantity using a grid of $75\times 75$ bins.}
\label{PR_poloidal_plane_30MeV}
\end{figure*}

We start our study of the collisionless pitch angle dispersion effects on synchrotron radiation emission by using mono-energetic and mono-pitch angle runaway electron distributions as the initial condition of KORC simulations. The kinetic energies (i.e. not including the rest mass energy $m_ec^2$) of the simulated runaways are $\mathcal{E}_0 = 10$ MeV and 30 MeV, and initial pitch angles of $\theta_0 = 5^\circ, 10^\circ, 15^\circ$, and $20^\circ$. This means that our initial distributions functions are delta functions in the energy and pitch angle $f_{RE}(\mathcal{E},\theta,t=0) = \delta (\theta - \theta_0)\delta (\mathcal{E} - \mathcal{E}_0)$.
The major radii of the torus used for the spatial initial condition are $R=1.475$ m and $R=1.43$ m for RE with $\mathcal{E} = 10$ MeV and 30 MeV, respectively. In all cases we use the radius of the RE beam $r=0.2$ m. In our simulations we evolve the runaway electrons by $t\sim 10 \ \mu$s, which is enough for reaching a steady-state distribution function.

In Fig.~\ref{PR_poloidal_plane_30MeV}(a) we show the spatial distribution of the total synchrotron radiated power $P_T$ of Eq.~(\ref{Ptot}) for the ensemble of runaway electrons with $\mathcal{E}=30$ MeV and $\theta_0=10^\circ$. The intensity of the radiation is higher at the high-field side (HFS) and lower at the low-field side (LFS), and the spatial distribution shows an up-down symmetry. Fig.~\ref{PR_poloidal_plane_30MeV}(b) shows the spatial distribution of the radiated power $P_R(\lambda)$ of Eq.~(\ref{P_lambda}) integrated over $\lambda\in (100,10000)$ nm. 
This range of wavelengths encompasses the visible and a part of the infrared portions of the electromagnetic spectrum, usually used in experimental studies.
The same qualitative features of $P_T$ are observed. 
Fig.~\ref{PR_poloidal_plane_30MeV}(c) shows the spatial distribution of runaways on the poloidal plane of the simulation of panels (a) and (b). For producing these figures we computed the histograms of each quantity using a grid of $75\times 75$ bins.
These same features of $P_T$ and the integrated synchrotron radiation power are observed in all the other simulations of initially mono-energetic and mono-pitch angle distributions of runaway electrons.

\begin{figure}[ht!]
\begin{center}
\includegraphics[scale=0.525]{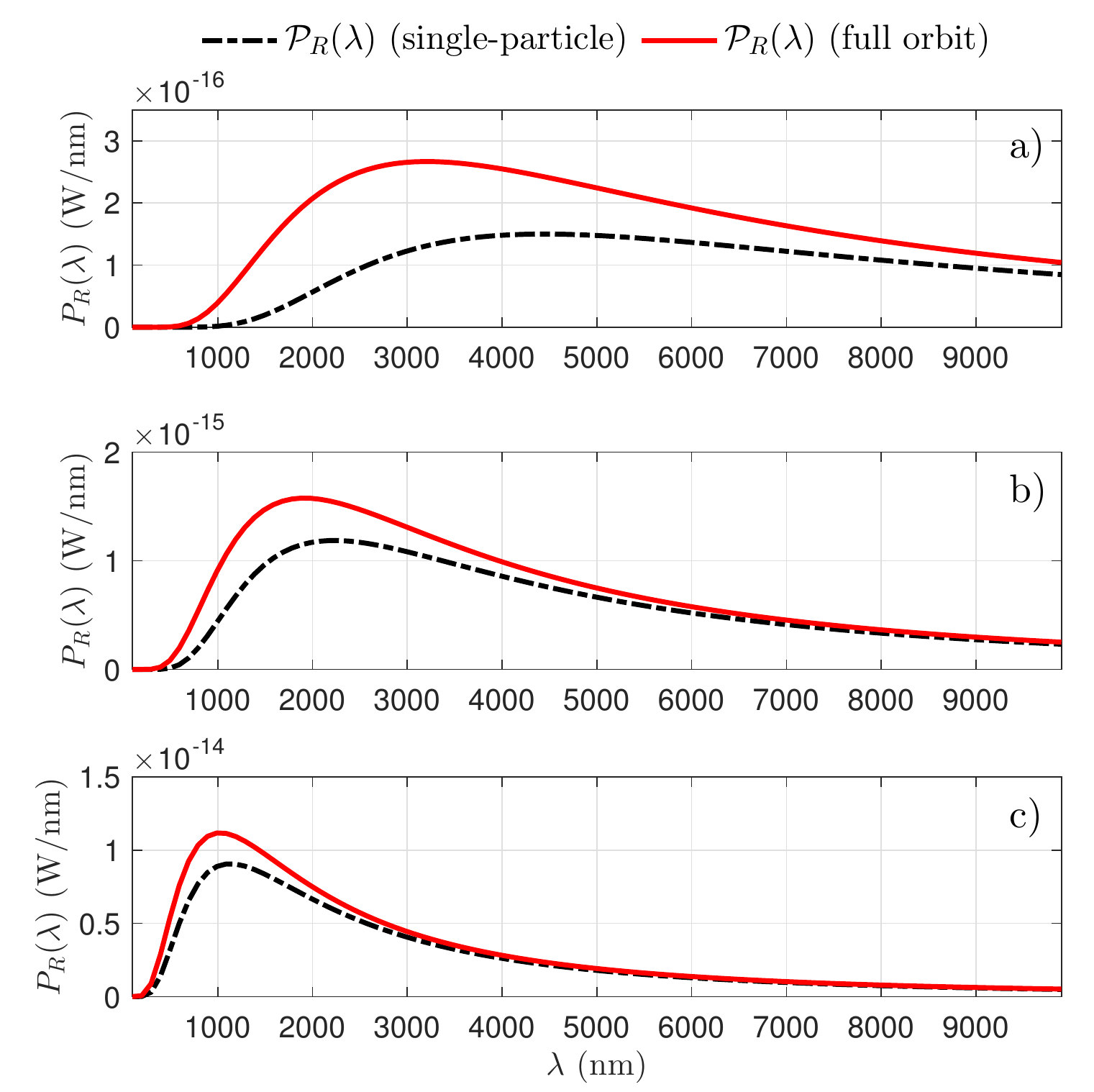}
\end{center}
\caption{Comparison between the synchrotron radiation spectra $\mathcal{P}_R(\lambda)$ and the corresponding single-particle spectra. Panel a): $\mathcal{P}_R(\lambda)$ in Eq.~(\ref{<PR>}) calculated for a simulation with $\mathcal{E}_0 = 30$ MeV and $\theta_0 = 5^\circ$. The corresponding single-particle spectrum is calculated using the above values for the energy and pitch angle and the value of the magnetic field at the magnetic axis $B = 2.1$ T. Panel b): same as panel a) for $\theta_0 = 10^\circ$. Panel c): same as panel a) for $\theta_0 = 20^\circ$.}
\label{spectra_poloidal_plane_30MeV}
\end{figure}
\begin{figure}[ht!]
\begin{center}
\includegraphics[scale=0.53]{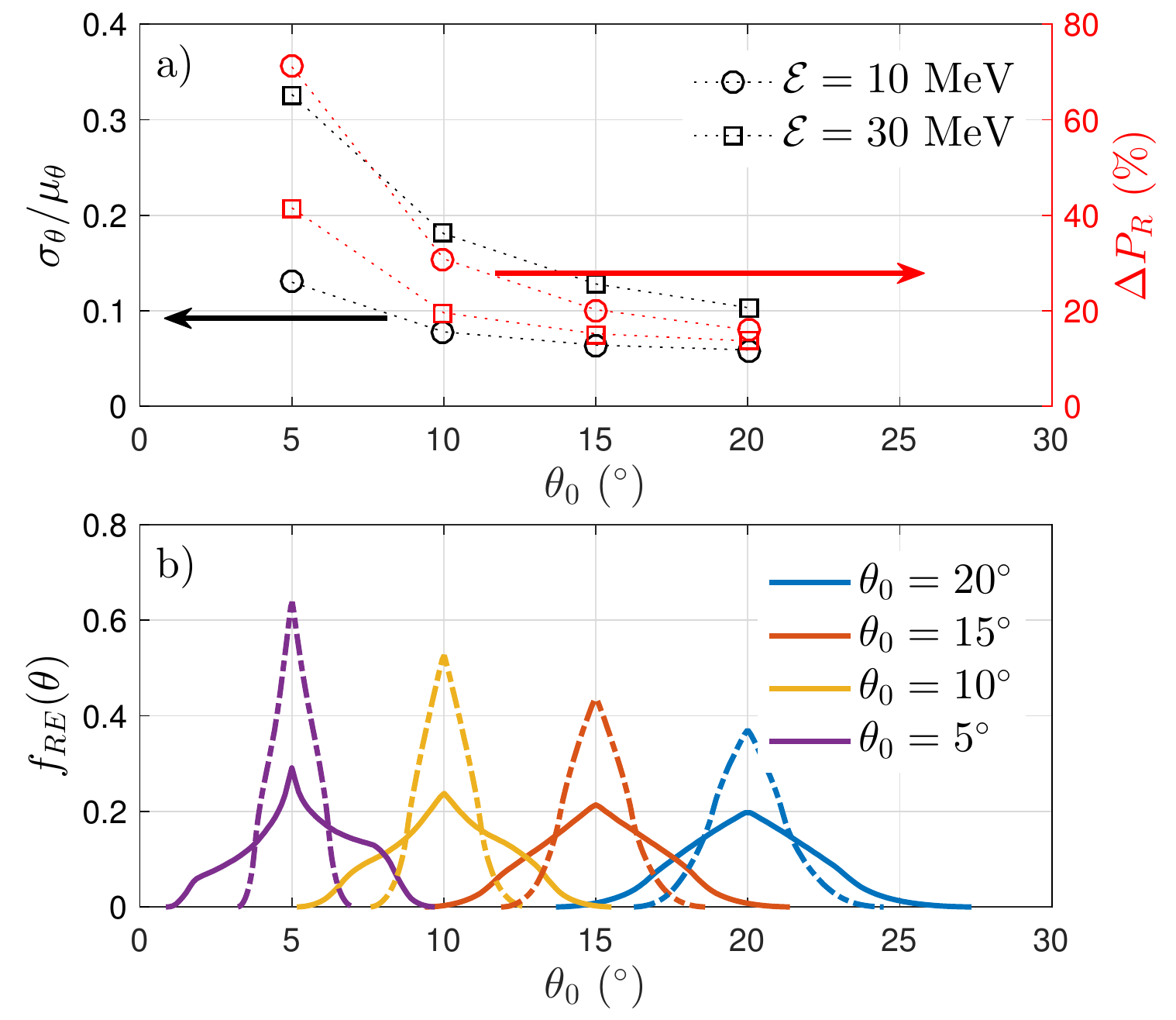}
\end{center}
\caption{Collisionless steady state distribution functions of runaway electrons. Panel a): left axis, relative dispersion of the pitch angle $\sigma_\theta/\mu_\theta$; right axis, the relative difference between the integrated power of the two spectra $\Delta P_R$ in the range of wavelengths $\lambda\in (100,10000)$ nm. Here $\mu_\theta$ and $\sigma_\theta$ are the mean and standard deviation of the full orbit $f_{RE}(\mathcal{E},\theta)$. Panel b): collisionless, steady state distribution functions of simulated runaway electrons for various initial pitch angles and the two energies $\mathcal{E}_0=10$ MeV (dashed lines), and $30$ MeV (solid lines). The departure of $\mathcal{P}_R(\lambda)$ from the single-particle spectra becomes larger as $\sigma_\theta/\mu_\theta$ becomes larger.}
\label{pitch_stats}
\end{figure}

In Fig.~\ref{spectra_poloidal_plane_30MeV} we show the comparison between the expected value of the synchrotron radiation spectra for different full orbit $f_{RE}(\mathcal{E},\theta)$, that is,

\begin{equation}
\mathcal{P}_R(\lambda) = \int \int  f_{RE}(\mathcal{E},\theta) P_R(\lambda,\mathcal{E},\theta) d \mathcal{E} d \theta\ ,
\label{<PR>}
\end{equation}

\noindent
and the so-called single-particle spectrum, namely, the synchrotron spectrum of Eq.~(\ref{P_lambda}) computed using the initial values for the energy and pitch angle of the runaways, and characteristic values for the magnetic field (taken at the magnetic axis). In this figure we only show the simulations with $\mathcal{E}_0 = 30$ MeV and $\theta_0=5^\circ,10^\circ,$ and $20^\circ$. The other simulations show similar results.
Among the differences between $\mathcal{P}_R(\lambda)$ and the corresponding single-particle spectra we observe that the maximum of $\mathcal{P}_R(\lambda)$ tends to move towards smaller wavelengths, and its magnitude is larger in all cases. These changes in the shape of $\mathcal{P}_R(\lambda)$ are particularly important because the runaway electrons' parameters are usually inferred by fitting the experimentally measured synchrotron spectrum with the single-particle spectrum.
In Ref.~\cite{Stahl2013} the authors used pre-computed distribution functions for the runaways to show that $\mathcal{P}_R(\lambda)$ can be very different from what is called the single-particle spectrum. This was also shown in Ref.~\cite{Landreman2014} for distribution functions of runaways obtained from solving the Fokker-Plank equation with radiation losses and collisions in 0-D simulations, that is, not including spatial information.
In our simulations any departure of $\mathcal{P}_R(\lambda)$ from the single-particle spectra results from allowing the magnetic field to have a spatial dependence, which in turn translated into collisionless pitch angle dispersion. In Fig.~\ref{pitch_stats}(b) we show the full orbit, steady state distribution functions. We observe that as the value of the relative dispersion of the pitch angle $\sigma_\theta/\mu_\theta$ (Fig.~\ref{pitch_stats}(a)) increases, the departure of $\mathcal{P}_R(\lambda)$ from the single-particle spectra becomes larger; we measure this departure using the relative difference between the integrated power of the two spectra in the range of wavelengths $\lambda\in (100,10000)$ nm, this is shown on the right axis of Fig.~\ref{pitch_stats}(a) as $\Delta P_R$.
Here $\mu_\theta$ and $\sigma_\theta$ are the mean and standard deviation of the full orbit $f_{RE}(\mathcal{E},\theta)$.

\subsection{Synchrotron emission of mono-energetic and mono-pitch angle RE distributions as measured by a camera}
\label{mono-section_camera}

\begin{figure*}[ht!]
\begin{center}
\includegraphics[scale=0.55]{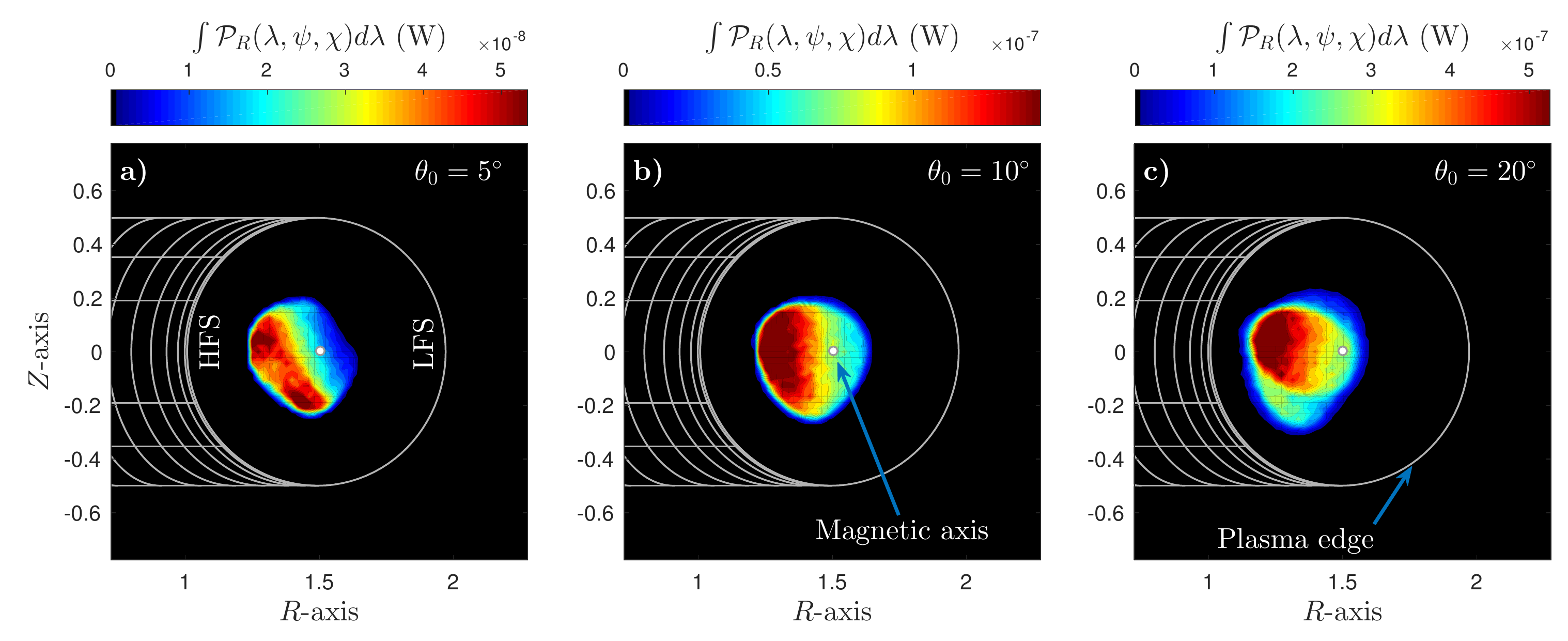}
\end{center}
\caption{Spatial distribution of the integrated synchrotron radiation of simulated runaway electrons with energy $\mathcal{E}_0 = 30$ MeV and various initial pitch angles as measured by a camera. Panel a): spatial distribution of the integrated synchrotron radiation for the simulation with initial pitch angle $\theta_0=5^\circ$. Panel b): same as panel a) for $\theta_0=10^\circ$. Panel c): same as panel a) for $\theta_0=20^\circ$. For this calculation the camera has been placed at the outer midplane plasma at a radial distance from the center of the plasma of $R_{sc}=2.4$ m. The other parameters of the camera are described in the appendix of Sec.~\ref{Apendix1}. For this calculation we have integrated the radiation over the range of wavelengths $\lambda \in(100,10000)$ nm. We observe a transition from a crescent shape to an ellipse shape for the spatial distribution of the radiation as we go from small to large initial pitch angles.}
\label{camera_30MeV}
\end{figure*}

We now go a step further and calculate the spatial distribution and spectra of the synchrotron radiation as measured by a camera placed at the outer midplane plasma. In this calculation each pixel of the camera measures the synchrotron radiation integrated along the corresponding line of sight.
To the best of our knowledge this calculation is the first of its kind, including the exact full-orbit dynamics of runaway electrons in toroidal magnetic fields and the basic geometric optics of a camera. 
In the Appendix we describe in detail the set-up of the camera in the simulations. For this calculation we have used the full orbit information of each electron in our simulations and two models for the angular distribution, namely, the full angular distribution $P_R(\lambda,\psi,\chi)$ in Eq.~(\ref{P_ang}) and the simplified model for the angular distribution $P_{R}^{\Omega_\alpha}(\lambda) = P_R(\lambda)/\Omega_\alpha$.
In Fig.~\ref{camera_30MeV} we show the spatial distribution of the integrated synchrotron emission of simulations with $\mathcal{E}_0=30$ MeV and $\theta_0=5^\circ, 10^\circ$, and $20^\circ$ calculated with $P_R(\lambda,\psi,\chi)$. We have integrated the radiation over the range of wavelengths $\lambda \in(100,10000)$ nm. No significant difference is observed if a visible or infrared filter is used for the synchrotron radiation. Using $P_R^{\Omega_\alpha}(\lambda)$ for calculating the spatial distribution of the synchrotron emission yields to qualitatively similar results, showing the same spatial features, but having an intensity one order of magnitude larger. 
Contrary to the spatial distribution of the synchrotron emission on the poloidal plane (c.f. Fig.~\ref{PR_poloidal_plane_30MeV}), the spatial distribution of the synchrotron emission seen by the camera shows a variety of different non-symmetric shapes, they transition from a crescent shape to an ellipse shape as the mean pitch angle increases. For distributions of runaway electrons with $\mathcal{E}_0 < 30$ MeV and with pitch angles in the range $\theta_0 \leq 20^\circ$ we always observe crescent shapes. 
In addition to the different shapes of the radiation seen by the camera, we observe a shift of the bright regions towards the HFS as we increase the pitch angle, despite the actual spatial distribution of the runaways remain fairly symmetric and localised around the magnetic axis, see Fig.~\ref{PR_poloidal_plane_30MeV}(c). This shift of the bright regions towards the HFS strongly depends on the pitch angle of the electrons, becoming larger as we increase the pitch angle; its dependence on energy is observed to be rather weak, increasing as we increase the energy only for $\theta_0\geq 20^\circ$.

\begin{figure}[ht!]
\begin{center}
\includegraphics[scale=0.525]{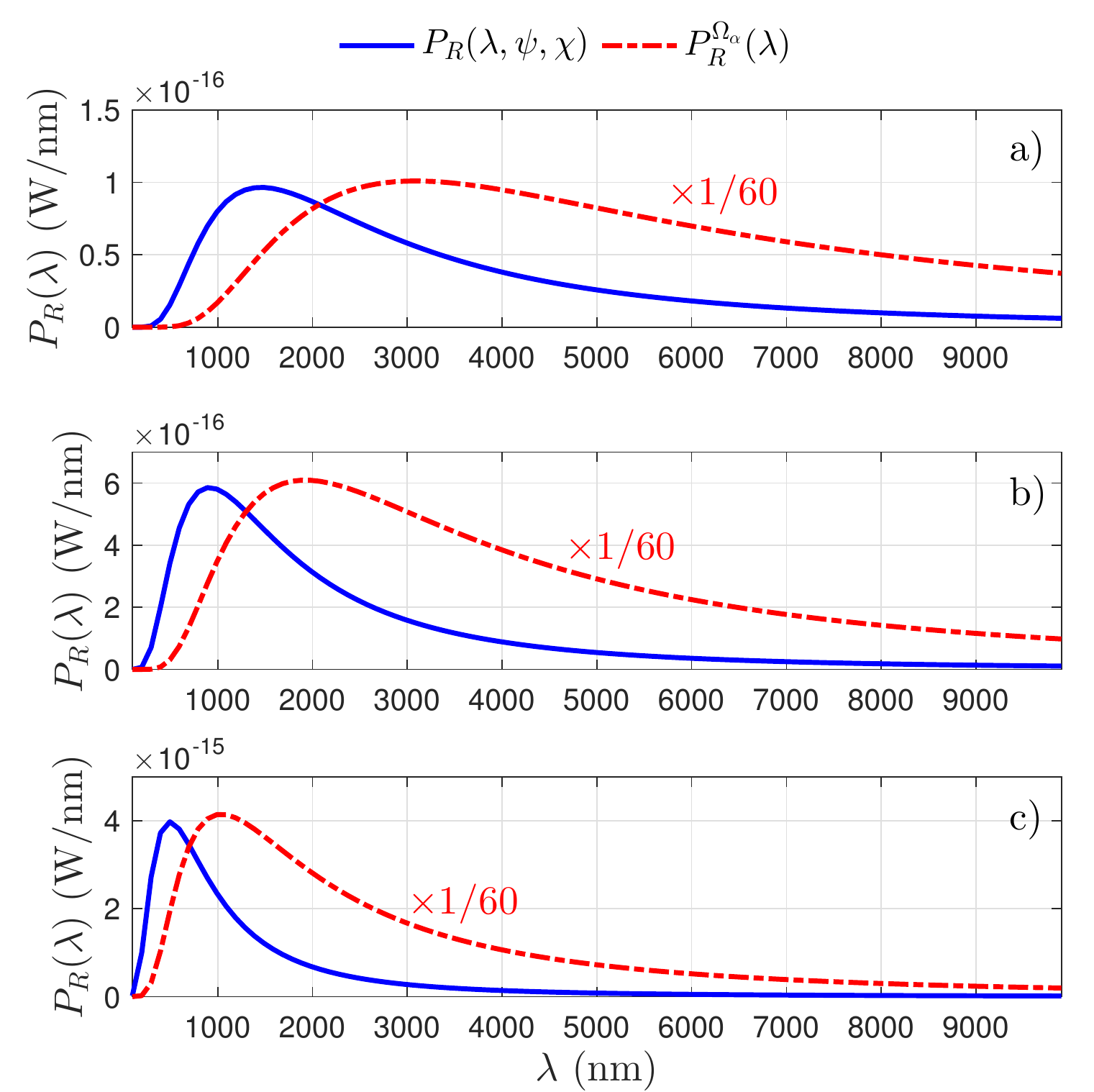}
\end{center}
\caption{Synchrotron radiation spectra of simulated runaway electrons with $\mathcal{E}_0=30$ MeV as measured by the camera. For comparison purposes we show the spectra calculated using $P_R(\lambda,\psi,\chi)$ (solid blue line) and $P_R^{\Omega_\alpha}(\lambda)$ (dashed red line). Panel a): synchrotron radiation spectra for an ensemble of runaways with $\theta_0=5^\circ$. Panel b): same as panel a) but for $\theta_0=10^\circ$. Panel c): same as panel a) but for $\theta_0=20^\circ$. The amplitude of the spectra is approximately sixty times larger when calculated using $P_R^{\Omega_\alpha}(\lambda)$ with respect to the result obtained with $P_R(\lambda,\psi,\chi)$. Also, the maximum of the spectra is shifted towards larger wavelengths when using $P_R^{\Omega_\alpha}(\lambda)$. These large differences may result in underestimating the runaway electron density and pitch angles of the runaway electrons if $P_R^{\Omega_\alpha}(\lambda)$ is used to interpret the experimental measurements.}
\label{spectra_camera_30MeV}
\end{figure}

Finally, we calculate the synchrotron radiation spectra of the simulated distributions of runaways as measured by the camera. In this case we regard the camera as one big spectrometer, merging the information of all the pixels of the camera. This calculation can be done using only one or a small subset of pixels of the camera if needed. 
In Fig.~\ref{spectra_camera_30MeV} we show the spectra of simulated runaway electrons with $\mathcal{E}_0 = 30$ MeV and various pitch angles. We calculate the spectra using both the full angular distribution $P_R(\lambda,\psi,\chi)$, and the simplified angular distribution $P_R^{\Omega_\alpha}(\lambda)$. 
The spectra calculated using the full angular distribution $P_R(\lambda,\psi,\chi)$ shows the same features than the spectra of Fig.~\ref{spectra_poloidal_plane_30MeV}, namely, the amplitude of the spectra becomes larger and the maximum of the spectra shifts towards smaller wavelengths as the pitch angle increases.
The differences between the spectra of $P_R(\lambda,\psi,\chi)$ and $P_R^{\Omega_\alpha}(\lambda)$ are in their magnitude, being approximately sixty times larger when calculated using $P_R^{\Omega_\alpha}(\lambda)$ than when using $P_R(\lambda,\psi,\chi)$, and in their shape, having the maximum of the spectra shifted towards larger wavelengths when using $P_R^{\Omega_\alpha}(\lambda)$; these large differences may result in underestimating the runaway electron density and pitch angles of the runaway electrons if $P_R^{\Omega_\alpha}(\lambda)$ is used to interpret the experimental measurements.
We have explored the case when the ``natural aperture'' $\alpha$ of the cone defining the emission region of $P_R^{\Omega_\alpha}(\lambda)$ becomes smaller than $1/\gamma$. In this case, the spatial distribution and the shape of the spectra of the synchrotron emission measured by the camera remains practically unchanged, but the amplitude of the synchrotron spectra becomes even larger than in the case where $\alpha=1/\gamma$.

\subsection{Synchrotron emission of avalanching RE on the poloidal plane}
\label{avalanching_runaways_poloidal}

Now we consider a more realistic distribution function for runaway electrons that might occur during the early times of a runaway disruption in tokamak plasmas, that is, the avalanche distribution function \cite{Rosenbluth1997,Fulop2006,Stahl2013}. This distribution function describes the exponential increase in time  of the runaway density during early times of a runaway disruption and is given by:

\begin{equation}
f_{RE}(p,\eta)  = \frac{\hat{E} p}{2\pi C_z \eta} \exp{\left( -\frac{p\eta}{C_z} - \frac{\hat{E}p}{2\eta}(1-\eta^2) \right)}\ ,
\label{avalanchePDF}
\end{equation}

\noindent
where $p = \gamma m_e v$ is the relativistic momentum of an electron, $\eta = \cos\theta$, $\hat{E} = (\bar{E}-1)/(1+Z_{eff})$, $Z_{eff}$ is the effective ion charge, $\bar{E} = E_\parallel/E_c$, $E_\parallel$ is the parallel electric field normalised to the critical electric field $E_c = m_ec/(e\tau_{coll})$, and $C_z=\sqrt{3(Z_{eff} + 5)/\pi} \log{\Lambda}$. 

\begin{figure}[ht!]
\begin{center}
\includegraphics[scale=0.54]{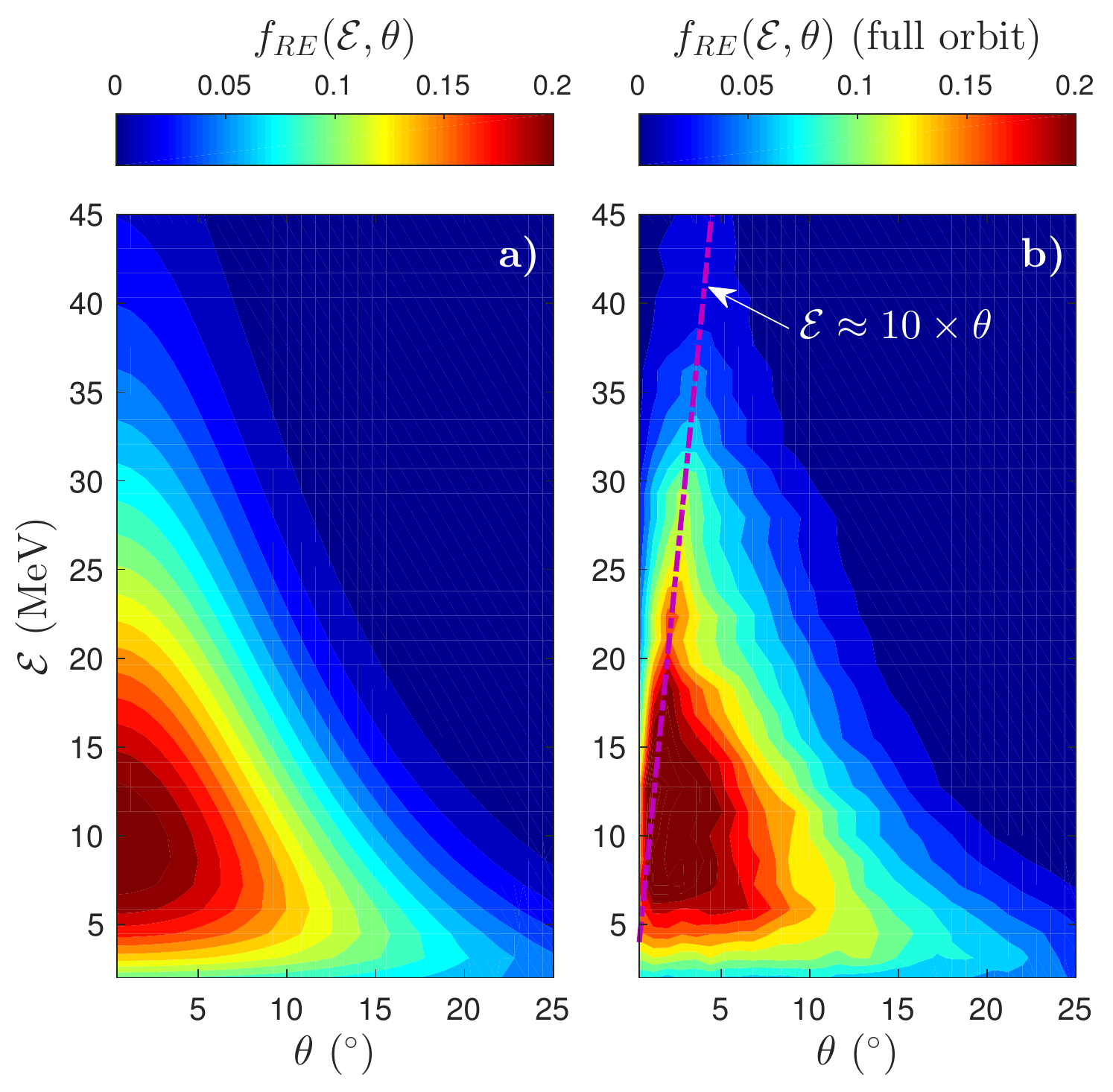}
\end{center}
\caption{Filled contours of the analytical and simulated avalanche distribution function for runaway electrons with $Z_{eff}=1$. Panel a): filled contours of the analytical $f_{RE}(\mathcal{E},\theta)$. Panel b): simulated distribution function by the end of the simulation. We infer a linear relation between the energy of the bulk of the distribution and the pitch angle given by $\mathcal{E} \approx 10\times \theta$.}
\label{avalanche_PDF}
\end{figure}

\begin{figure}[ht!]
\begin{center}
\includegraphics[scale=0.525]{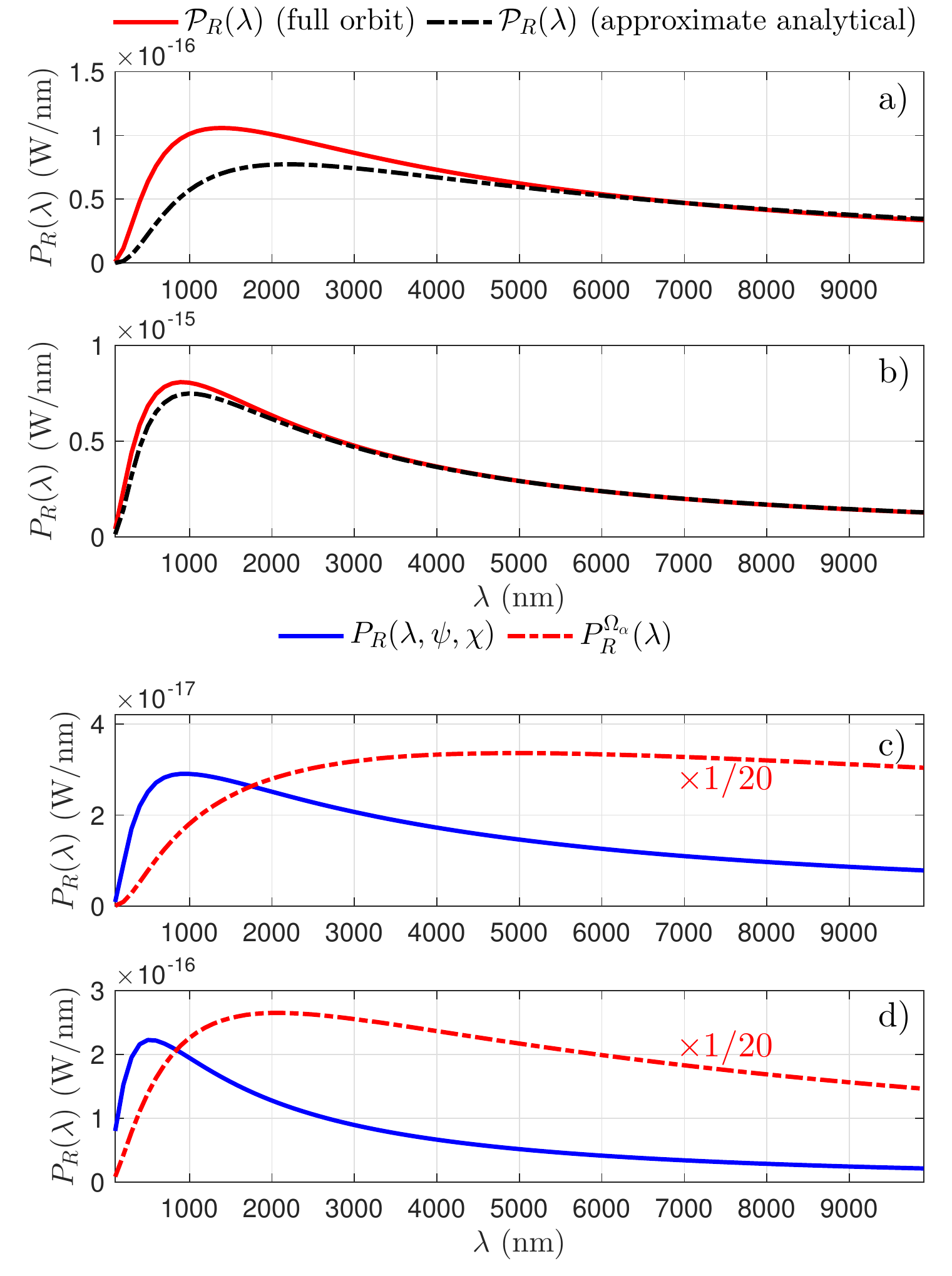}
\end{center}
\caption{Expected value of the synchrotron radiation spectra of simulated avalanche distribution functions for runaway electrons (a)-(b), and synchrotron radiation spectra as measured by a camera placed at the outer midplane plasma (c)-(d). Panel a): synchrotron radiation spectra of Eq.~(\ref{<PR>}) (solid red line) for the avalanche distribution function with $Z_{eff}=1$. The dashed black line shows the approximate analytical spectra using directly Eq.~(\ref{avalanchePDF}) into Eq.~(\ref{<PR>}). Panel b): same as panel a) for $Z_{eff}=10$. Panel c): synchrotron radiation spectra as measured by the camera for the case $Z_{eff}=1$. Panel d): same as panel c) for the case with $Z_{eff}=10$.}
\label{spectra_avalanche}
\end{figure}

We use Eq.~(\ref{avalanchePDF}) as the initial condition of our simulations with $n_e=3.9\times10^{20}$ m$^{-3}$, which results in $\tau_{coll} \sim 10$ ms, $E_c=0.15$ V/m, and we consider $Z_{eff}=1$ and $Z_{eff}=10$ for simulating an hydrogenic plasma and a plasma with high concentration of impurities, respectively. 
We use $E_\parallel = 0.74$ V/m so that it is in agreement with typical values of the loop voltage measured in DIII-D plasmas during runaway disruptions~\cite{Stahl2013,Hollmann2013,Yu2013}. Larger (smaller) values of $E_\parallel$ result in narrower (wider) pitch-angle distributions and longer (shorter) tails of the energy distribution of avalanching runaways. Therefore different values of $E_\parallel$ leading to different avalanche distributions modify the corresponding synchrotron emission.
The major radius of the torus used for the spatial initial condition is $R=1.37$ m, and the radius of the RE beam is set to $r=0.2$ m. In Fig.~\ref{avalanche_PDF}(a) we show the filled contours of $f_{RE}(\mathcal{E},\theta)$ using Eq.~(\ref{avalanchePDF}) with $Z_{eff}=1$; using $Z_{eff}=10$ results in a wider distribution in pitch angle space at low energies $\mathcal{E}\sim 10$ MeV. Here, $\mathcal{E} = c\sqrt{p^2 + m_e^2 c^2}$ and $\theta=\arccos{\eta}$. We observe only small fluctuations for the difference between the analytical and the initial condition of our simulations, that is, $\sqrt{(f_{RE}-f_{sim})^2 }\sim 0.01$, where $f_{sim}$ is the sampled distribution function used as the initial condition of our simulations. We sample $f_{RE}(p,\eta)$ using the Metropolis-Hastings algorithm. By the end of the simulations $f_{RE}(\mathcal{E},\theta)$ have reached a steady state, in Fig.~\ref{avalanche_PDF}(b) we show the simulated distribution function which shows departures from the initial condition, specially at large energies $\mathcal{E}\geq 20$ MeV. We infer a linear relation between the energy of the bulk of the distribution and the pitch angle given by $\mathcal{E} \approx 10\times \theta$. 

\begin{figure*}[ht!]
\begin{center}
\includegraphics[scale=0.55]{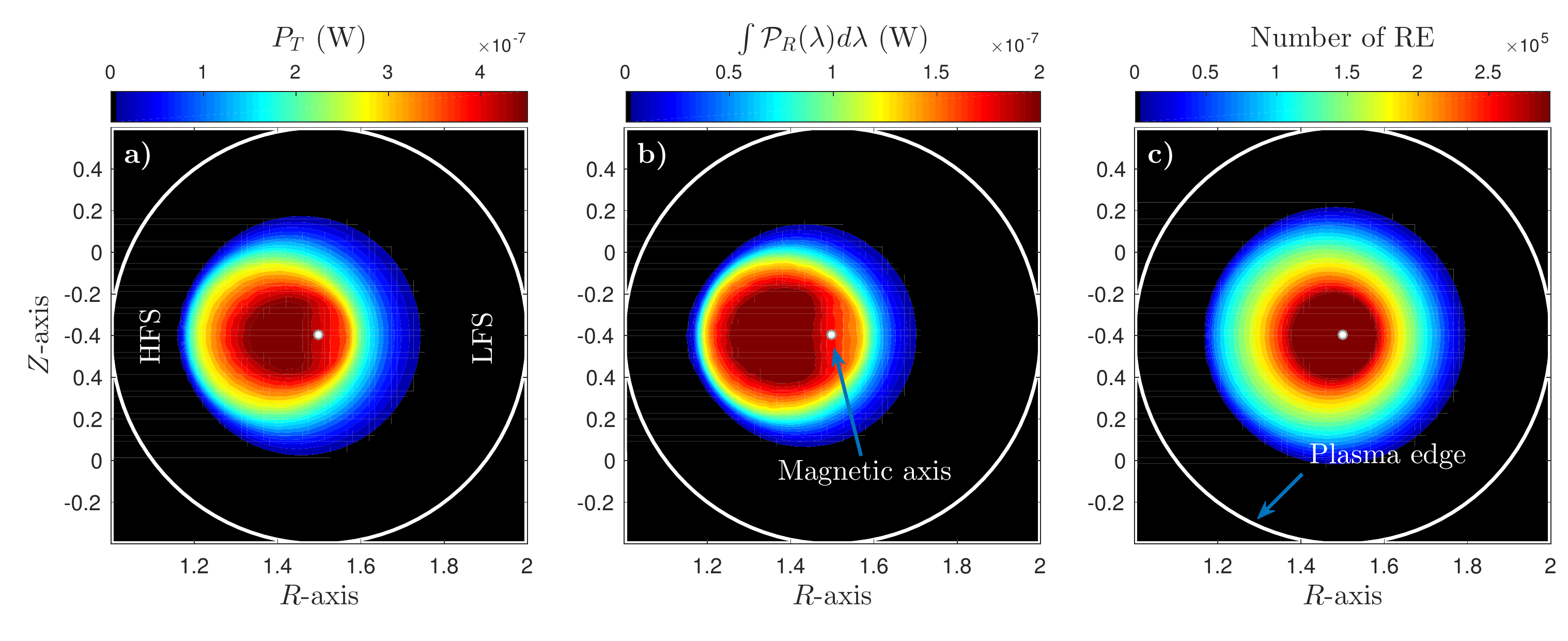}
\end{center}
\caption{Spatial distribution of the total and integrated synchrotron radiated power of the simulated avalanche distribution function for runaway electrons by the end of the simulation.
Panel a): spatial distribution of the total synchrotron radiated power $P_T$ of Eq.~(\ref{Ptot}). Panel b): spatial distribution of the radiated power $P_R(\lambda)$ of Eq.~(\ref{P_lambda}) integrated over $\lambda\in (100,10000)$ nm. Panel c): the spatial distribution of the simulated runaways by the end of the simulation. Notice that the bright regions of radiation not necessarily corresponds to the more dense regions. For producing these figures we computed the histograms of each quantity using a grid of $75\times 75$ bins.}
\label{PR_poloidal_plane_Zeff1}
\end{figure*}

As for the case of the mono-energy and mono-pitch angle distributions, we first calculate the spatial distribution of the total and the integrated synchrotron radiated power for the avalanche distribution function. This is shown in Fig.~\ref{PR_poloidal_plane_Zeff1} for the case with $Z_{eff}=1$, we obtain the same qualitative results for $Z_{eff}=10$. This time, the spatial distribution on the poloidal plane of $P_T$ and the integrated $P_R(\lambda)$ shows more structure, with a bright region of radiation with a crescent shape at the HFS. Notice that the bright regions of radiation not necessarily corresponds to the more dense regions, see Fig.~\ref{PR_poloidal_plane_Zeff1}(c).

In Fig.~\ref{spectra_avalanche}(a)-(b) we show (red solid line) the spectra $\mathcal{P}_R(\lambda)$ in Eq.~(\ref{<PR>}) of the simulated avalanche distribution functions with $Z_{eff}=1$ and $Z_{eff}=10$, respectively. We also show for comparison the spectra computed directly using Eq.~(\ref{avalanchePDF}) (dashed black line). As it can be seen, the spectra of the simulated avalanche distributions show the same trends as the mono-energy and mono-pitch angle distributions: a larger amplitude, and the shift of the maxima of $\mathcal{P}_R(\lambda)$ towards smaller wavelengths. However, as we increase $Z_{eff}$ the differences between the approximate analytical and the full orbit $\mathcal{P}_R(\lambda)$ become smaller. 

\subsection{Synchrotron emission of avalanching RE as measured by a camera}
\label{avalanching_runaways_camera}

Next, we compute the spatial distribution and the spectra of the synchrotron radiation as measured by a camera placed a the outer midplane plasma. For this calculations the parameters of the camera are the same as in Sec.~\ref{mono-section_camera} and in the appendix. In Fig.~\ref{camera_avalanche} we show the spatial distribution of the integrated synchrotron radiation calculated using the full angular distribution $P_R(\lambda,\psi,\chi)$. We have integrated the radiation over the range of wavelengths $\lambda \in(100,10000)$ nm. No significant difference is observed if a visible or infrared filter is used for the synchrotron radiation. Using the simplified angular distribution $P_R^{\Omega_\alpha}(\lambda)$ results in similar features of the spatial distribution of the radiation. 
Consistent with the results of Sec.~\ref{mono-section_camera}, we observe the transition from a crescent to an ellipse shape for the spatial distribution of the radiation as we increase $Z_{eff}$, as we are effectively increasing the pitch angle of the bulk of the runaway distribution function.

The crescent shape of the spatial distribution of the synchrotron radiation observed in Fig.~\ref{camera_avalanche}(a) and \ref{camera_30MeV}(a) results from the contribution of runaway electrons with small pitch angle that follow the winding of the magnetic field lines.
In Fig.~\ref{camera_avalanche_sections} we show the contribution of different toroidal sectors of the runaway beam to Fig.~\ref{camera_avalanche}(a); as it can be seen, the larger contribution to the crescent shape of the synchrotron radiation spatial distribution comes from the toroidal sector with $\varphi \in (40^\circ,70^\circ)$, where $\varphi$ is the toroidal angle as defined in Fig.~\ref{camera_setup}(c).
As the pitch angle of the runaways increases, their velocity vector is not longer pointing along the magnetic field lines, resulting in shapes similar to Fig.~\ref{camera_avalanche}(b) and \ref{camera_30MeV}(c).

Finally, we calculate the spectra of the synchrotron radiation as measured by the camera, these are shown in Fig.~\ref{spectra_avalanche}(c)-(d). As for the simulations of Sec.~\ref{mono-section_camera}, we observe large differences between the spectra calculated using the two different angular distributions for the radiation, namely, the magnitude of the spectra calculated using $P_R^{\Omega_\alpha}(\lambda)$ is approximately twenty times larger than when using $P_R(\lambda,\psi,\chi)$, also the maximum of the spectra are shifted towards larger wavelengths in the case when $P_R^{\Omega_\alpha}(\lambda)$ is used. As discussed before, this may result in underestimating the runaway electron density and pitch angles of the runaway electrons if $P_R^{\Omega_\alpha}(\lambda)$ is used to interpret the experimental measurements.

\begin{figure}[ht!]
\begin{center}
\includegraphics[scale=0.53]{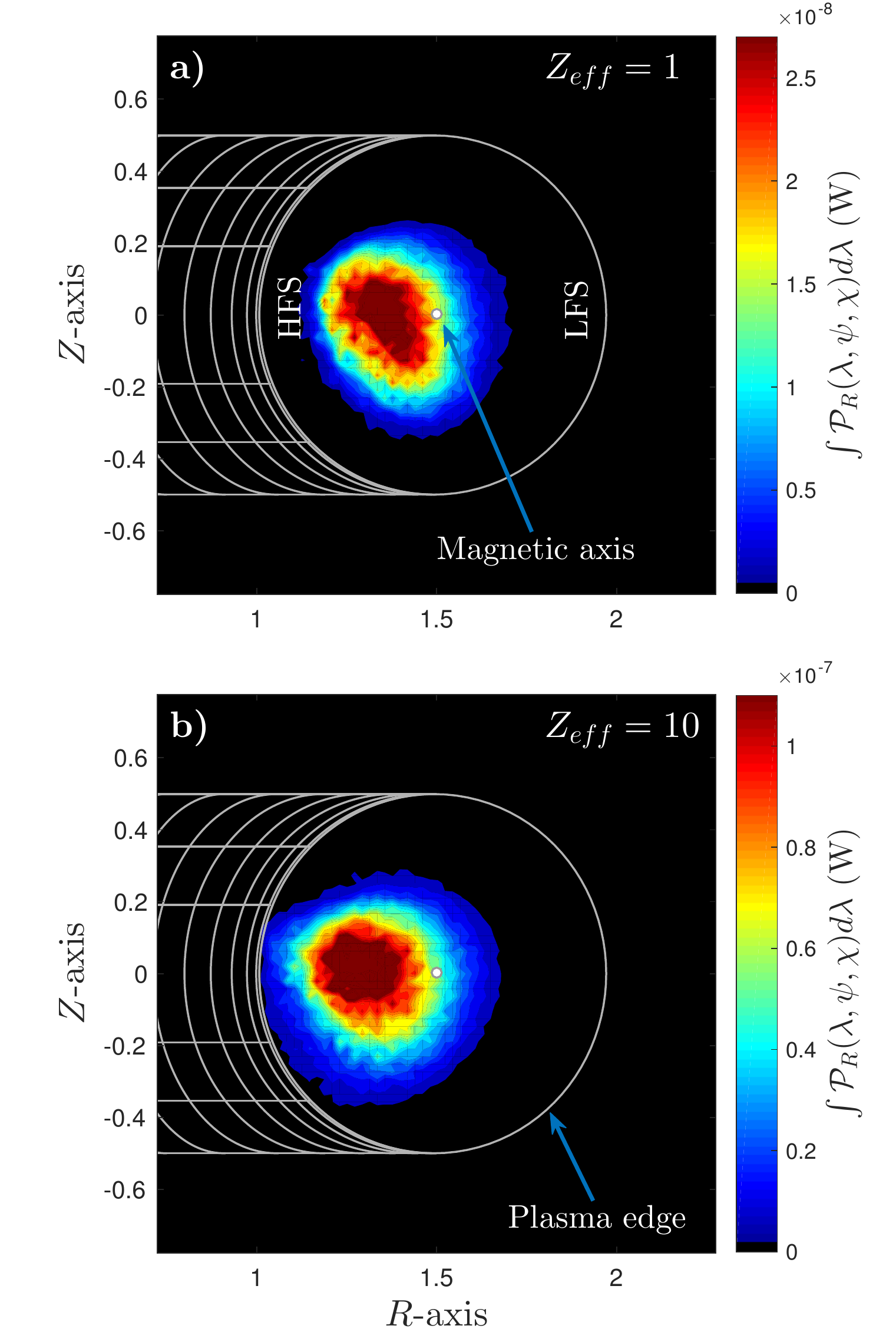}
\end{center}
\caption{Spatial distribution of the integrated synchrotron radiation of the full orbit avalanche distribution function for runaway electrons as measured by a camera. Panel a): spatial distribution of the integrated synchrotron radiation for the avalanche distribution with $Z_{eff}=1$. For this calculation we have used the full angular distribution $P_R(\lambda,\psi,\chi)$. Panel b): same as panel a) for $Z_{eff}=10$. The parameters of the camera are the same as in Fig.~\ref{camera_30MeV}. We observe a transition from a crescent shape to an ellipse shape for the bright regions of the radiation as we go from small to large values of $Z_{eff}$.}
\label{camera_avalanche}
\end{figure}

\section{Discussion and conclusions}
\label{conclusions}

In this paper we have addressed the long standing question about what are the relationships between different  runaway electrons distribution functions and their corresponding synchrotron emission including: full-orbit effects, information of the spectral and angular distribution of synchrotron radiation of each electron, and the basic geometric optics of a camera. 
We performed kinetic simulations of the full-orbit dynamics of different ensembles of runaway electrons in DIII-D-like magnetic fields to study in detail various aspects of their synchrotron emission. 

In Sec.~\ref{mono-section_poloidal} and \ref{mono-section_camera}, we used mono-energetic and mono-pitch angle distribution functions as the initial conditions of the simulations. For these simulations we calculated the spatial distribution on the poloidal plane of the total and the integrated synchrotron radiated power, which show bright regions of radiation at the HFS and up-down symmetry. Then we compared the synchrotron spectra of the full orbit distributions of runaways with the so-called single-particle spectra, showing that full orbit effects and in particular collisionless pitch angle dispersion effects cause the former to depart from the single-particle spectra. These effects become more evident as the relative dispersion of the pitch angle $\sigma_\theta/\mu_\theta$ increases, see Fig.~\ref{spectra_poloidal_plane_30MeV} and \ref{pitch_stats}. 

Then, we calculated the spatial distribution and spectra of the synchrotron radiation as measured by a camera placed at the outer midplane plasma. To the best of our knowledge this calculation is the first of its kind, including the exact full-orbit dynamics of runaway electrons in toroidal magnetic fields and the basic geometric optics of a camera. 
We used two models for the angular distribution of the synchrotron radiation, namely, the full spectral and angular distribution of Eq.~(\ref{P_ang}), and a simplified model where the radiation is emitted isotropically within a circular cone with ``natural aperture'' $\alpha = 1/\gamma$. 
Using either model for the angular distribution we observed a rich variety of non-symmetric shapes for the spatial distribution of the radiation that strongly depend on the pitch angle distribution of the runaways, and weakly depend on the runaways energy distribution, value of the $q$-profile at the plasma edge, and the chosen range of wavelengths.
We noticed a transition from a crescent shape to an ellipse shape as the mean pitch angle increases, see Fig.~\ref{camera_30MeV}. On the other hand, we found that the magnitude of the synchrotron spectra measured by the camera is overestimated by approximately a factor of 60 when the angular distribution is oversimplified, and the shape is affected too, moving to larger wavelengths when we use the simplified angular distribution $P_R^{\Omega_\alpha}(\lambda)$, see Fig.~\ref{spectra_camera_30MeV}. This may result in underestimating the runaway electron density and pitch angles of the runaway electrons if $P_R^{\Omega_\alpha}(\lambda)$ is used to interpret the experimental measurements.

In Sec.~\ref{avalanching_runaways_poloidal} and \ref{avalanching_runaways_camera} we repeated the analysis of previous sections for an avalanche RE  distribution function. We studied the case of a hydrogenic plasma ($Z_{eff}=1$) and a plasma with a high content of impurities ($Z_{eff}=10$). 
We find that collisionless pitch angle dispersion modifies the initial distribution function (c.f. Fig.~\ref{avalanche_PDF}), so that there exist a deviation of the pitch angle of the bulk distribution as function of the runaways' energy, that is, $\mathcal{E} \approx 10\times \theta$.
In this case we also observed a complex structure of the spatial distribution of the synchrotron radiation on the poloidal plane with a non-trival relation to the spatial density of runaway electrons, see Fig.~\ref{PR_poloidal_plane_Zeff1}. As in the simulations of Sec.~\ref{mono-section_poloidal}, the synchrotron spectra of the full orbit avalanche distributions depart from the analytical approximation, showing larger departures for the case of $Z_{eff}=1$, see Fig.~\ref{spectra_avalanche}(a)-(b).
On the other hand, the spatial distribution of the synchrotron emission measured by the camera in our simulations showed a transition from a crescent shape to an ellipse shape as we increased $Z_{eff}$, this due to the effective increase of the pitch angle of the bulk distribution as $Z_{eff}$ becomes larger, c.f. Fig.~\ref{camera_30MeV}. 
We expect that in longer time scales, especially in plasmas containing high-Z impurities, the collisionless pitch-angle dispersion will be modified by collisions. This is a problem that we plan to address in a future publication.
Regarding the synchrotron spectra measured by the camera, {\bf similarly as in the simulations of Sec.~\ref{mono-section_camera}, we found that its amplitude is overestimated by approximately a factor of 20 when $P_R^{\Omega_\alpha}(\lambda)$ is used}, and its maximum is shifted to larger wavelengths with respect to the spectra of Eq.~(\ref{P_ang}), see Fig.~\ref{spectra_avalanche}(c)-(d).

The results reported in this paper show a weak dependence with the value of the q-profile at the plasma edge, remaining qualitatively and quantitatively similar. A more detailed analysis for investigating the dependence of the runaways synchrotron emission with different shapes of the q-profile is not in the scope of the present study.

These results shed some light into the relationship between a given runaway distribution function and its corresponding synchrotron emission in magnetic confinement plasmas. This might help to find better ways to interpret experimental measurements of synchrotron radiation to obtain better estimates of the runaway electron parameters, and so help to both formulate better theoretical descriptions of the runaways in these plasmas, and to improve the mechanisms for avoiding and/or mitigating runaway electrons.

\begin{figure*}[!ht]
\begin{center}
\includegraphics[scale=0.55]{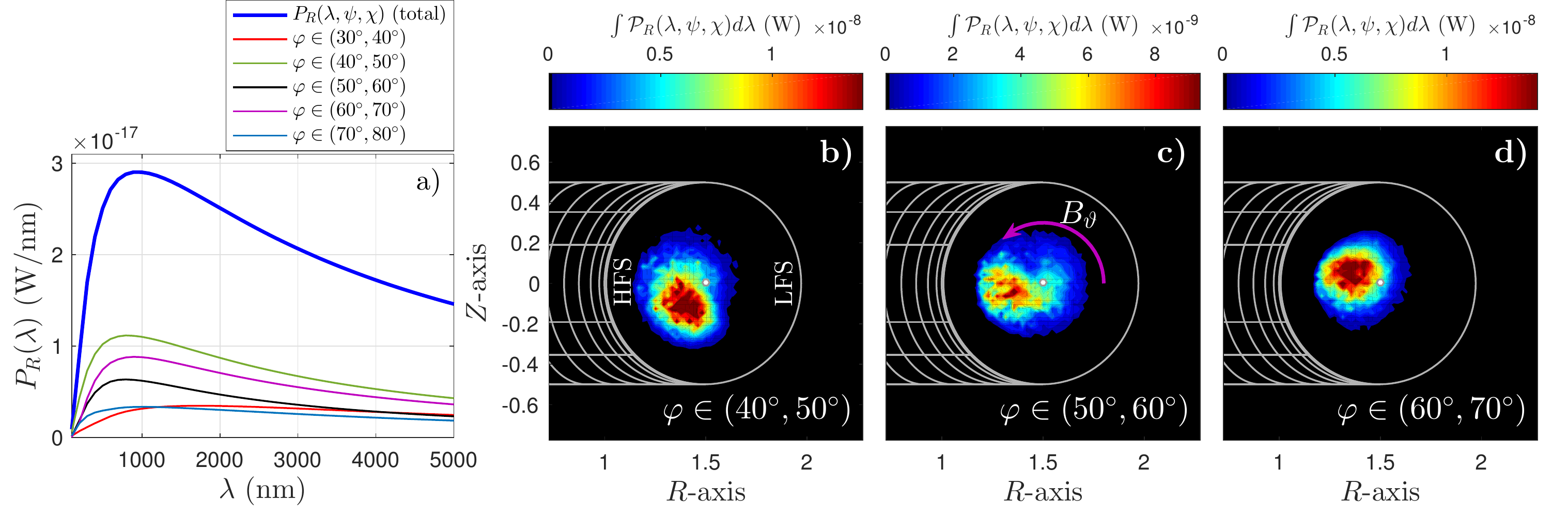}
\end{center}
\caption{Contribution of different toroidal sectors to the spatial distribution of the integrated synchrotron radiation of the full orbit runaways avalanche distribution function with $Z_{eff}=1$. Panel a): contribution of some toroidal sectors to the synchrotron spectra of Fig.~\ref{spectra_avalanche}(c). Panels b) to d): synchrotron radiation as measured by the camera of the toroidal sectors $\varphi\in(40^\circ,50^\circ)$, $\varphi\in(50^\circ,60^\circ)$, and $\varphi\in(60^\circ,70^\circ)$, respectively.
The crescent shape of the spatial distribution of the synchrotron radiation observed in Fig.~\ref{camera_avalanche}(a) results from the contribution of runaways  with small pitch angle that follow the winding of the magnetic field lines.}
\label{camera_avalanche_sections}
\end{figure*}

\section{Acknowledgments}
This material is based upon work supported by the U.S. Department of Energy, Office of Science, Office of Fusion Energy Sciences under Contract No. DE-AC05-00OR22725.
Research sponsored by the Laboratory Directed Research and Development Program of Oak Ridge National Laboratory, managed by UT-Battelle, LLC, for the U. S. Department of Energy.
This research used resources of the National Energy Research Scientific Computing Center, a DOE Office of Science User Facility supported by the Office of Science of the U.S. Department of Energy under Contract No. DE-AC02-05CH11231.

\section{Appendix: Setup of the synthetic camera diagnostic}
\label{Apendix1}

The camera diagnostic in KORC consists of an array of pixels on a rectangular detector placed at $\bm{R}_{sc} = (R_{sc},Z_{sc})$, where in a cylindrical coordinate system with origin at the center of the tokamak, $R_{sc}$ is the cylindrical radial position of the camera, and $Z_{sc}$ the corresponding camera position along the $z$-axis. 
For simplicity we assume that the camera is placed along the $x$-axis of a Cartesian coordinate system, and that the radial camera position $R_{sc}$ defines the outer wall at the midplane plasma, too.
The horizontal and vertical size of the camera detector determine the optics of the camera, that is, the horizontal and vertical angles of view of the camera. 
We assume that the camera has a single lens located at $\bm{R}_{sc}$, so that each pixels has a single line of sight that connects the center of each pixel to the center of the lens, and then extends into the plasma.

\begin{figure*}[ht!]
\begin{center}
\includegraphics[scale=0.575]{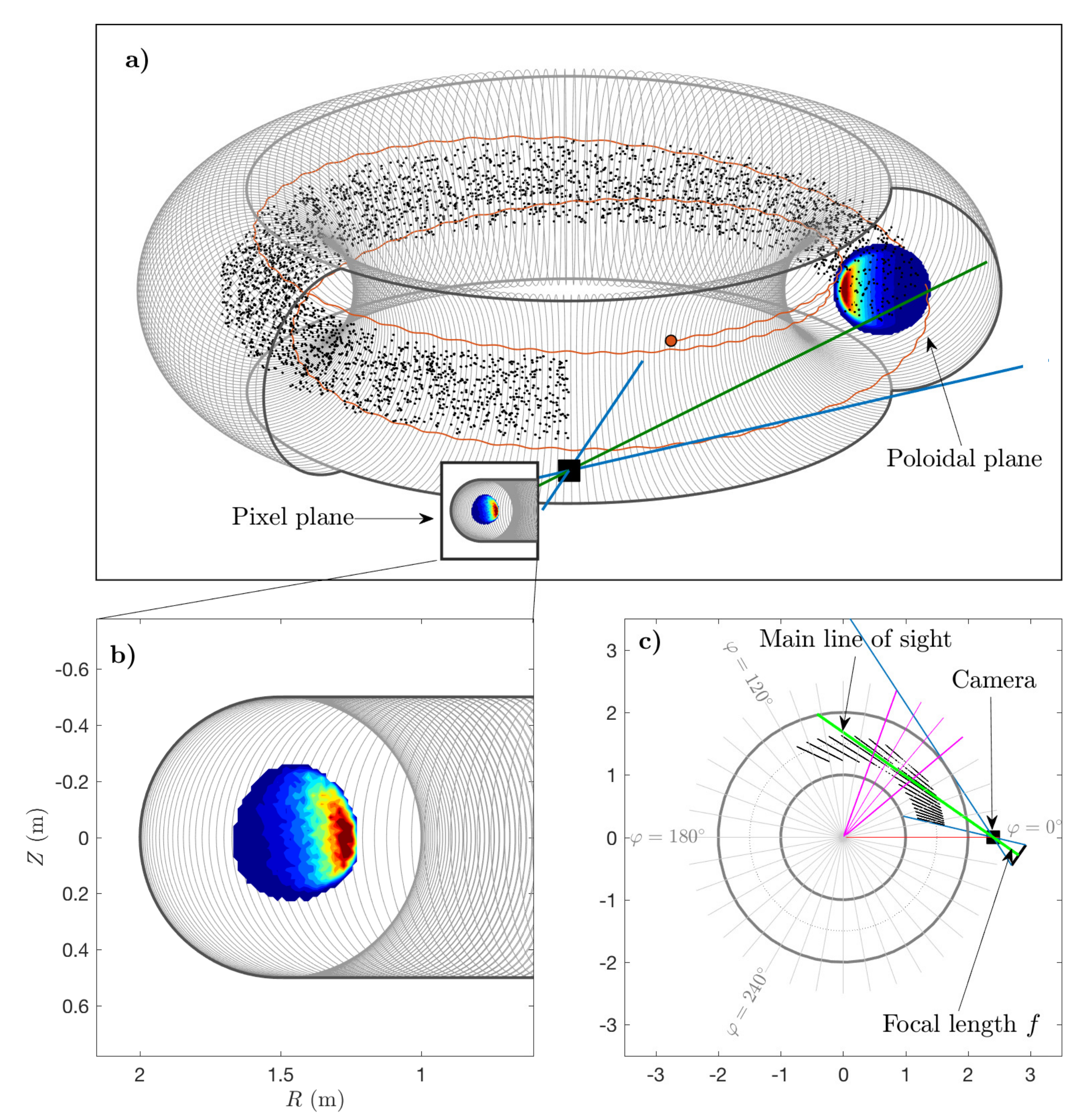}
\end{center}
\caption{Camera setup in KORC simulations. Panel a): schematic representation of the camera setup showing the horizontal angle of view of the camera (blue lines), the main line of sight of the camera (green line), the position of the camera (black square), the synchrotron emission at the poloidal plane and at the detector plane (a.k.a. pixel plane), and an example of the initial spatial distribution of the simulated runaway electrons (black dots).
Panel b): zoom of the detector plane of the camera showing an example of the measured synchrotron emission in a KORC simulation. 
Panel c): top view of the camera setup showing some examples of lines of sight of the camera in a KORC simulation. The toroidal sectors used in Fig.~\ref{camera_avalanche_sections} are highlighted in magenta.}
\label{camera_setup}
\end{figure*}

In Fig.~\ref{camera_setup}(a) we show the setup of the synthetic camera placed at $R_{sc} = 2.4$ and $Z_{sc} =0$. 
The size of the detector is $40\ \mbox{cm}\times 40\ \mbox{cm}$, and the pixel array is made of $75 \times 75$ pixels.
The blue lines show the horizontal angle of view, while the green line shows the main line of sight, that is, the line of sight joining the center of the detector and the lens. In Fig.~\ref{camera_setup}(c) we show the top view of the camera setup. In this figure the dotted lines show some lines of sight of different pixels of the camera.
Another parameter of the camera is its focal length $f$, which is the distance between the lens and the center of the camera detector and is chosen to be $f=50$ cm.
Finally, the incline of the camera $\vartheta$, which is the angle between the main line of sight (green line) and the solid horizontal red line in Fig.~\ref{camera_setup}(c), can be used to aim the camera. We choose $\vartheta = 55^\circ$ for all the simulations in this work.
In this way, the size of the detector, the focal length of the camera, and the incline of the camera determines the camera's field of view (See Fig.~\ref{camera_setup}(b)).

The frequency at which the camera can take snapshots is equal to or lower than the inverse of the time step used in a KORC simulation, with an exposure time that depends on how many snapshots are used to produce the final picture of the synchrotron radiation. 
Each pixel of the camera measures the line integrated synchrotron emission over the whole exposure time.

On the other hand, in axisymmetric plasmas the electromagnetic fields and particles' variables are independent of the cylindrical azimuthal angle $\phi$ (or the toroidal angle $\zeta$ in toroidal coordinates). Thus, any rigid rotation of the electron's variables by an arbitrary angle in the azimuthal direction is a possible realization of an electron in the plasma. The above implies that an electron can be detected by more than one pixel of the camera. In the camera set-up of Fig.~\ref{camera_setup} the azimuthal angle $\phi$ is measured anticlockwise.
A potential complication is that for every snapshot taken by the camera the radiation spectra would have to be calculated for each electron of the simulation--in a typical KORC simulation we simultaneously follow hundreds of thousands ($\sim 10^5$) of runaway electrons. It can be seen that for an array of $100\times100$ pixels the number of computations involved is larger than $10^9$, increasing quadratically with the number of pixels of the camera detector. 
This computation can become computationally costly if it is not done in an efficient way. In order to reduce the number of computations involved, we pre-select those runaways that are more likely to be seen by the camera.
The pre-selection of the electrons is done as follows: for each electron with velocity $\bm{v}_i$ and position $\bm{R}_i$, we extend $\bm{v}_i$ and calculate $\bm{R}_i^*=(R_{sc},Z_i^*)$, the point at which $\bm{v}_i$ intersects the outer wall. Here, the outer wall is modeled as an infinitely long cylindrical shell with inner radius $R_{sc}$. Note that $\bm{v}_i$ is a vector with origin at the electron position $\bm{R}_i$. 
Then, we measure the angle between the electron's velocity and the vector $\bm{R}^*_i - \bm{R}_i$, which is given by $\cos{\varsigma_i} = \bm{v}_i \cdot (\bm{R}^*_i - \bm{R}_i)/|\bm{v}_i| |\bm{R}^*_i - \bm{R}_i|$. 
In the simplest approximation for the angular distribution of the synchrotron radiation, the radiation is emitted within a circular cone with its axis along $\bm{v}_i$, and aperture $\alpha = 1/\gamma$, where $\gamma$ is the relativistic gamma factor of the particle. See Sec.~\ref{theory} for details. Only electrons with $\varsigma _i\leq \alpha$ are kept for the calculation of the camera snapshot.
Next, we iterate over each pixel of the camera detector and calculate the contribution of each electron to the line integrated emission measured by that pixel.
The process for computing the radiation emitted by the $i$-th electron and measured by each pixel is a two-step process: The first step is to find the columns of pixels that detect the $i$-th electron. We note that the pixels in the same column of the camera detector share the same line of sight when the camera setup is seen from the top, see Fig.~\ref{camera_setup}(c).
We say that the $i$-th electron is detected by the $j$-th column of pixels when the circle with radius $R_i = \sqrt{x_i^2 + y_i^2}$ defined by the position of the electron intersects the line of sight of that column of pixels. Here $\bm{x}_i = (x_i,y_i,z_i)$ is the position of the $i$-th electron. For the $i$-th electron seen by the $j$-th column of pixels we calculate the angle $\varphi_{i,j}$, which is the angle between the camera position and the position at which the circle with radius $R_i$ intersects the $j$-th line of sight. This angle is measured anticlockwise from the solid red line of Fig.~\ref{camera_setup}(a).
In the second step we identify the row of pixels that detect the $i$-th electron. This is done by identifying the row of pixels that the unitary vector $\hat{\bm{n}}_i$, the direction of emission of the $i$-th electron, hits when it extends from the electrons' position to the plane of the camera detector. Here  $\hat{\bm{n}}_i = \hat{T}_{-\varphi_{i,j}}\hat{T}_{\phi_0}\bm{x}_i - \hat{\bm{R}}_{sc}$, $\hat{T}_\varphi$ are rigid rotations along the $z$-axis by an angle $\varphi$, and $\phi_0$ is the azimuthal angle defined by the position of the particle $\bm{x}_i$.
Once that we have identified which pixels detect which electrons we compute their contribution to the measured synchrotron emission using either model for the angular distribution of the synchrotron radiation of Sec.~\ref{theory}.

\bibliography{Carbajal_PPCF_2017}

\end{document}